%% file: proceedings.tex
\documentclass[preprint]{sigchi}

\setcopyright{acmcopyright}
\doi{}
\isbn{}

\usepackage{balance}       
\usepackage{graphics}      
\usepackage[T1]{fontenc}   
\usepackage{txfonts}
\usepackage{mathptmx}
\usepackage[pdflang={en-US},pdftex]{hyperref}
\usepackage{color}
\usepackage{booktabs}
\usepackage{textcomp}
\usepackage[utf8]{inputenc}
\usepackage{amssymb}

\usepackage{microtype}        
\usepackage{ccicons}          

\usepackage{todonotes}
\usepackage{subcaption}
\usepackage{enumitem}

\def\plaintitle{A Visual Programming Paradigm for Abstract Deep Learning Model Development}
\def\plainauthor{Srikanth G Tamilselvam, Naveen Panwar, Shreya Khare,
  Rahul , Rahul Aralikatte, Anush Sankaran, Senthil Mani \\ IBM Research}

\def\plainkeywords{Deep Learning, Visual Programming, drag-and-drop interface, auto programming}

\makeatletter
\def\url@leostyle{%
  \@ifundefined{selectfont}{
    \def\UrlFont{\sf}
  }{
    \def\UrlFont{\small\bf\ttfamily}
  }}
\def\@copyrightspace{\relax}
\makeatother
\def\pprw{8.5in}
\def\pprh{11in}

\definecolor{linkColor}{RGB}{6,125,233}
\hypersetup{%
  pdftitle={\plaintitle},
 pdfauthor={\plainauthor},
  pdfkeywords={\plainkeywords},
  pdfdisplaydoctitle=true, 
  bookmarksnumbered,
  pdfstartview={FitH},
  colorlinks,
  citecolor=black,
  filecolor=black,
  linkcolor=black,
  urlcolor=linkColor,
  breaklinks=true,
  hypertexnames=false
}

\graphicspath{{./images/}}

\begin{document}

\title{\plaintitle}

\author{Srikanth Tamilselvam,\textsuperscript{†}
Naveen Panwar,\textsuperscript{†}
Shreya Khare,\textsuperscript{†}\\
Rahul Aralikatte,\textsuperscript{‡}
Anush Sankaran,\textsuperscript{†}
Senthil Mani,\textsuperscript{†}\\
\textsuperscript{†}{IBM Research},
\textsuperscript{‡}{University of Copenhagen}}

\maketitle
\begin{abstract}

Deep learning is one of the fastest growing technologies in computer science with a plethora of applications. But this unprecedented growth has so far been limited to the consumption of deep learning experts. The primary challenge being a steep learning curve for learning the programming libraries and the lack of intuitive systems enabling non-experts to consume deep learning. Towards this goal, we study the effectiveness of a ``no-code" paradigm for designing deep learning models. Particularly, a visual drag-and-drop interface is found more efficient when compared with the traditional programming and alternative visual programming paradigms. 
We conduct user studies of different expertise levels to measure the entry level barrier and the developer load across different programming paradigms. 
We obtain a System Usability Scale (SUS) of $90$ and a NASA Task Load index (TLX) score of $21$ for the proposed visual programming compared to $68$ and $52$, respectively, for the traditional programming methods.

\end{abstract}



\input{1_introduction}

\input{2_related_works}
\input{3_system_design}

\input{4_user_study}
\input{5_study_results}

\input{7_design_implication}

\input{8_conclusion}

\bibliographystyle{SIGCHI-Reference-Format}
\bibliography{ref}

\end{document}

%% file: 1_introduction.tex
\section{Introduction} 
\label{introduction}


Deep learning (DL) is one of the most pervasive and breakthrough technologies of the previous decade~\cite{sejnowski2018deep}. Automation has progressed significantly in various application domains such as self-driving cars \cite{angelova2015real}~\cite{bojarski2016end}~\cite{huval2015empirical}, flight control systems \cite{zhang2016learning}, and medical diagnosis~\cite{milletari2016v}~\cite{esteva2017dermatologist} primarily due to the advent of deep learning technology. According to Gartner, $80\%$ of the data scientists will have DL in their toolkits by 2018~\footnote{\url{https://www.gartner.com/newsroom/id/3804363}}. With this sudden explosion of DL, there is a requirement for current software engineers and data scientists to learn and upskill themselves in a short span of time. However, DL model authoring has been a skill that is currently restricted only to experts~\footnote{\url{https://www.kdnuggets.com/2018/01/democratizing-ai-deep-learning-machine-learning-dell-emc.html}}. The democratization of DL model development has been inhibited due to the following challenges:
\begin{enumerate}
\item \textbf{High Initial Learning Curve:} There is a steep learning curve involved in understanding the fundamentals of deep learning and the syntax nuances of multiple deep learning authoring libraries. There is a lack of intuitive systems to author DL models in an easy fashion. 
\item \textbf{Lack of Interaction across Libraries:} Different programming libraries exist for DL implementation such as Keras\footnote{\url{https://keras.io/}}, Tensorflow\footnote{\url{https://www.tensorflow.org/}}, PyTorch\footnote{\url{https://pytorch.org/}}, Caffe\footnote{\url{https://caffe2.ai/}}. However, there is limited interoperability across these libraries that enables a model designed and trained in one specific library to be accessible and re-usable across libraries. 
\item \textbf{Theoretical Knowledge:} Theory of deep learning has a strong mathematical prior. Additionally, the ability of making the right architecture choices and hyper-parameter preferences comes with practice and experience.
\end{enumerate}

\noindent Sankaran et al.~\cite{Sankaran:2017} studied the challenges faced by a DL developer by conducting a qualitative survey among $100$ software engineers from varying backgrounds. $83\%$ of the participants responded that it took them about $3-4$ days to implement a deep learning model, given the model design and choice to use any DL library. Interestingly, $86\%$ of those respondents had rated themselves as highest in ``programming ability". Thus, even good programmers find it challenging to implement and author deep learning models. This motivates the need for an efficient and intuitive method for developing deep learning models. The survey also showed that more than $92\%$ of the respondents wanted an interoperability framework to convert the code and model implemented in one library into another library. In the recent past, a few open source unified libraries were made available in the community such as Open Neural Network Exchange (ONNX)~\cite{ONNX71:online}, \textit{nGraph} from Intel~\cite{Intel:online}, and Neural Network Exchange Format (NNEF)~\cite{NeuralNe42:online}. These unified frameworks provide a standard, library agnostics representation of deep learning models. Despite the availability of such standardized frameworks, the ease in adoption of DL for non-expert users and rapid prototyping for expert users still remains a challenge.  

\begin{table*}[]
\centering
\resizebox{\textwidth}{!}{%
\begin{tabular}{|l|l|l|l|l|l|l|}
\hline
\multicolumn{1}{|c|}{\textbf{Tool}} & \multicolumn{1}{c|}{\textbf{\begin{tabular}[c]{@{}c@{}}UI \\ Designing\end{tabular}}} & \multicolumn{1}{c|}{\textbf{\begin{tabular}[c]{@{}c@{}}Model \\ Visualization\end{tabular}}} & \multicolumn{1}{c|}{\textbf{\begin{tabular}[c]{@{}c@{}}Code \\ Available\end{tabular}}} & \multicolumn{1}{c|}{\textbf{\begin{tabular}[c]{@{}c@{}}Multi-library \\ support\end{tabular}}} & \multicolumn{1}{c|}{\textbf{\begin{tabular}[c]{@{}c@{}}Training \\ Dashboard\end{tabular}}} & \multicolumn{1}{c|}{\textbf{\begin{tabular}[c]{@{}c@{}}Inference \\ API\end{tabular}}} \\ \hline
Weka~\cite{frank16:_weka_workb} & Limited & Limited & Yes & No & Yes & Yes \\ \hline
Digits~\cite{NVIDIADI52:online} & No & Yes & No & No & Yes & No \\ \hline
Aetros~\cite{HomeAETR26:online} & Yes & No & Yes & No & Yes & No \\ \hline
Fabrik~\cite{cloudCVF0:online} & Yes & Yes & Yes & Yes & No & No \\ \hline
Tensorboard~\cite{TensorBo51:online} & No & Yes & No & No & Yes & No \\ \hline
Azure ML & No & No & No & No & Yes & No \\ \hline
NN Console~\cite{NeuralNe37:online} & Yes & Yes & No & No & Yes & No \\ \hline
Activis~\cite{DBLP:journals/corr/KahngAKC17} & No & Yes & No & No & No & No \\ \hline
Netron~\cite{lutzroed65:online} & No & Yes & No & No & No & No \\ \hline
Deep Cognition~\cite{DeepCogn92:online} & Yes & Yes & No & No & Yes & Yes \\ \hline
Machine UI~\cite{MachineU79:online} & Yes & Yes & Yes & No & Yes & No \\ \hline
DL-IDE (proposed) & Yes & Yes & Yes & Yes & Yes & Yes \\ \hline
\end{tabular}%
}
\caption{Comparison of different tools and frameworks in the literature that enables easy and quick development of deep learning models.}
\label{tbl:related_work}
\end{table*}

In this research, we develop an easy to use visual programming paradigm and an Integrated Development Environment (IDE) for intuitive designing and authoring of deep learning models. This DL-IDE aims to democratize deep learning model development for software engineers and data scientists by offering a ``no-code" designing platform and reducing the learning curve. 

Fundamentally, any deep learning model design can be visualized as a graph data structure with a collection of nodes as computational layers and each layer having a list of hyper-parameters. To ease designing of such models, the DL-IDE has a drag-and-drop framework to visually design the deep learning model. The layer connections and the layer parameters could be set through the user interface minimizing the need for coding. DL-IDE represents the designed deep learning model as an abstract representation called Neural Language Definition Standard (NLDS) enabling execution ready code to be generated in multiple libraries such as Tensorflow, Keras, PyTorch, and Caffe. To study the effectiveness of the proposed DL-IDE, we conducted usability study among $18$ intermediate and expert deep learning developers, comparing the visual method and programming method for designing deep learning models.  The primary aim of the user study is to verify the hypothesis that, in comparison with the traditional coding method for designing deep learning models, the visual programming~\cite{myers1990taxonomies} method is better in terms of ease-of-use, ease in adoption, and decreases the prototyping efforts. The key research contributions can be summarized as follows:
\begin{enumerate}
    \item \textbf{Visual DL-IDE:} A visual programming IDE enabling ``no-code" intuitive way of designing deep learning models\footnote{To abide to the double blind format of the paper, only screenshots of the system is made available. The entire system will be made publicly available upon the acceptance of the paper.}.
    \item \textbf{Neural Language Definition Standard (NLDS):} An abstract representation capturing all the required parameters of a DL model, independent of the underlying implementation library (Tensorflow, PyTorch etc.) or language. 
    \item \textbf{Static Model Validation:} An extensive collection of manually curated rules to automatically validate the design parameters of a deep learning model in real-time, performed on top of NLDS.
    \item \textbf{Multiple Library Code Generation:} Adopting Model Driven Development of generating bug free, execution ready code in multiple libraries from NLDS representation. Currently, we support Tensorflow v1.4, Keras v2.1, PyTorch v0.3, and Caffe v1.0. 
    \item \textbf{System Usability Study:} A user study is conducted with $18$ intermediate and expert deep lear   ning developers to assess, and compare the traditional programming method with the proposed visual drag-and-drop user interface. 
\end{enumerate}

The rest of the paper is organized as follows: Section 2 provides a literature survey comparing the DL-IDE with other abstract visual programming tools. Section 3 details the design goals and the system architecture of the proposed deep learning IDE. Section 4 explains the user study setup and the data collection process. Section 5 analyzes the experimental results from the user study. Section 6 discusses the design implications of our study setup, and Section 7 concludes this research work discussing some potential future extensions.


%% file: 2_related_works.tex
\section{Related Work}
\label{lit_study}

The main requirements for any deep learning tool to be usable can be summarized into three major categories:

\begin{itemize}
\item \textbf{Design and Visualization}: This includes features such as intuitive designing and construction of deep learning models, visualization of pre-designed models, and dashboarding of model logs during training.
\item \textbf{Elimination of need to code}: This includes features such as automatic source code generation, support for multiple libraries, and import from existing codebases.
\item \textbf{Production-ready capabilities}: This includes features such as efficiently training the model, support for easy inference, and fast and scalable deployment.
\end{itemize}

\noindent Most of today's tools are aimed at addressing one or at most two of these categories. Table~\ref{tbl:related_work} summarizes the various tools features. The concept of easy designing of models dates back to Weka~\cite{frank16:_weka_workb}. Weka was ``no-code" method for designing traditional machine learning tasks having limited support for neural networks. It provides a dashboard for visualizing training metrics and a Java-based API for inference. NVIDIA DIGITS~\cite{NVIDIADI52:online} was the first deep learning specific tool to provide an user interface to perform tasks like data loading and one-click training. However, it is only possible to train pre-built networks and it does not have the capability to design models from scratch. 
More recently, tools like Google's Tensorboard~\cite{TensorBo51:online} is often used as a supplement to visualize Tensorflow models and Netron~\cite{lutzroed65:online} is used to visualize existing models coded in various libraries. 

Though these visualization tools help, there is still a need to visually author a new model from scratch and not just visualize an existing model. There is a need for an end-to-end system aimed to seamlessly integrating the design and training procedures. Sony's Neural Network Console~\cite{NeuralNe37:online}, Aetros~\cite{HomeAETR26:online}, Deep Cognition~\cite{DeepCogn92:online}, Fabrik~\cite{cloudCVF0:online}, and Machine UI~\cite{MachineU79:online} provide an intuitive drag-and-drop based interface to design custom deep learning models. They also have a compute backend to train the designed models and present the training metrics in a real-time dashboard. Although, new models could be designed using the drag-and-drop interface, most of these do not allow the corresponding execution ready code to be downloaded. These tools use their own compute backend for training but users often refuse to use such tools as uploading confidential data into these third party services is conflicting.

Thus it can be observed that, though there exists multiple tools which perform one or more tasks sufficiently, there is a requirement to address the problem as a whole with an end-to-end perspective. Therefore, we created DL-IDE which supports visual authoring, generation of code in multiple existing libraries, and download design (NLDS) as well as code for DL model development.

%% file: 3_system_design.tex
\section{Deep Learning IDE (DL-IDE) System Architecture}
\label{system}

\begin{figure}
    \begin{subfigure}[b]{0.45\textwidth}            \includegraphics[width=3.2in]{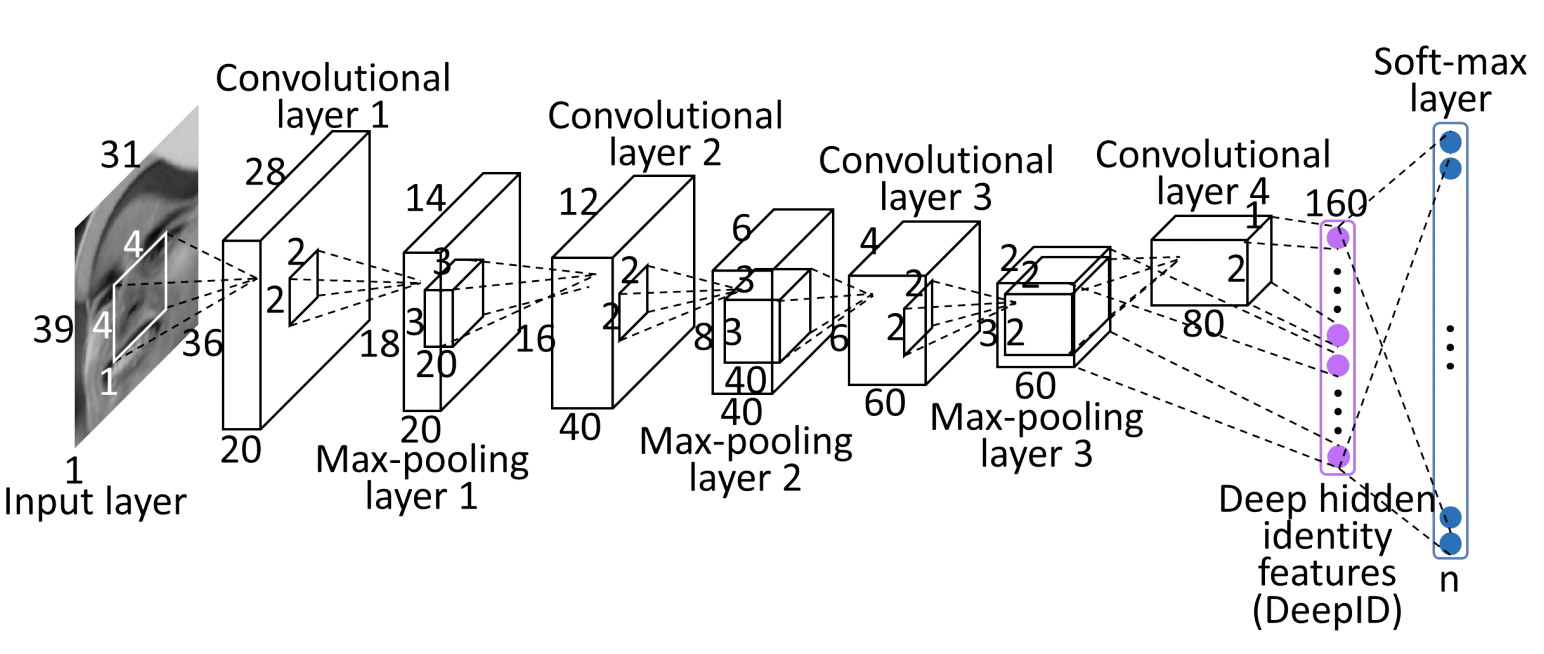}
        \caption{Flow figure~\cite{sun2014deep}}
        \label{fig:figure}
    \end{subfigure}
    \begin{subfigure}[b]{0.45\textwidth}
        \includegraphics[width=3.2in]{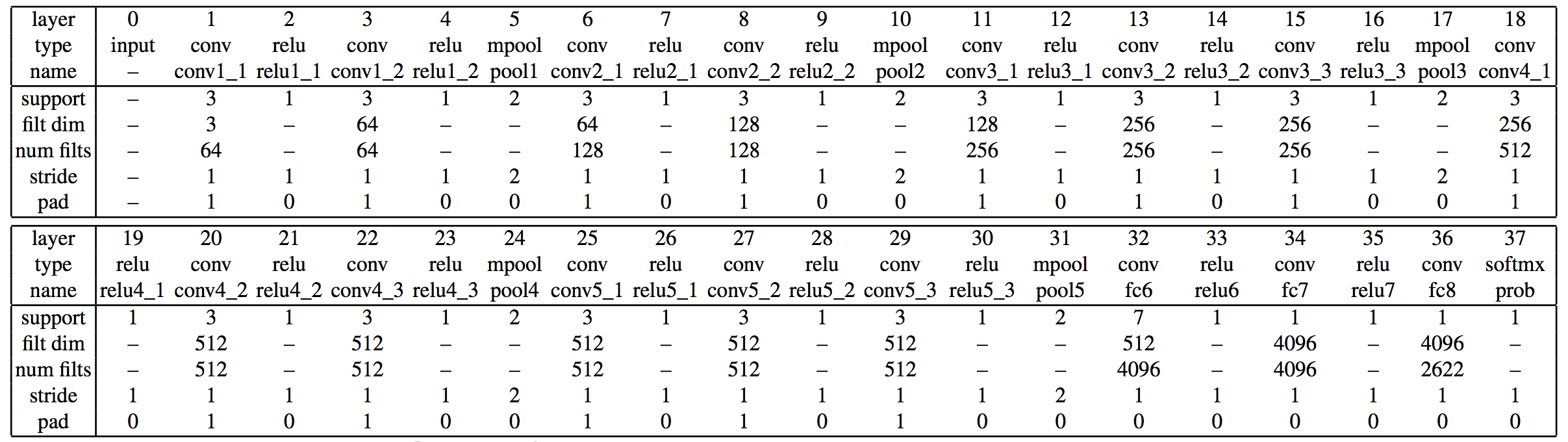}
        \caption{Tabular format~\cite{parkhi2015deep}}
        \label{fig:table}
    \end{subfigure}
    \caption{Two popular methods for expressing deep learning model design in research papers, presentations, and documents. Developers typically implement the models by following these representations.}
\end{figure}

\begin{figure*}
\centering
\includegraphics[width=\textwidth]{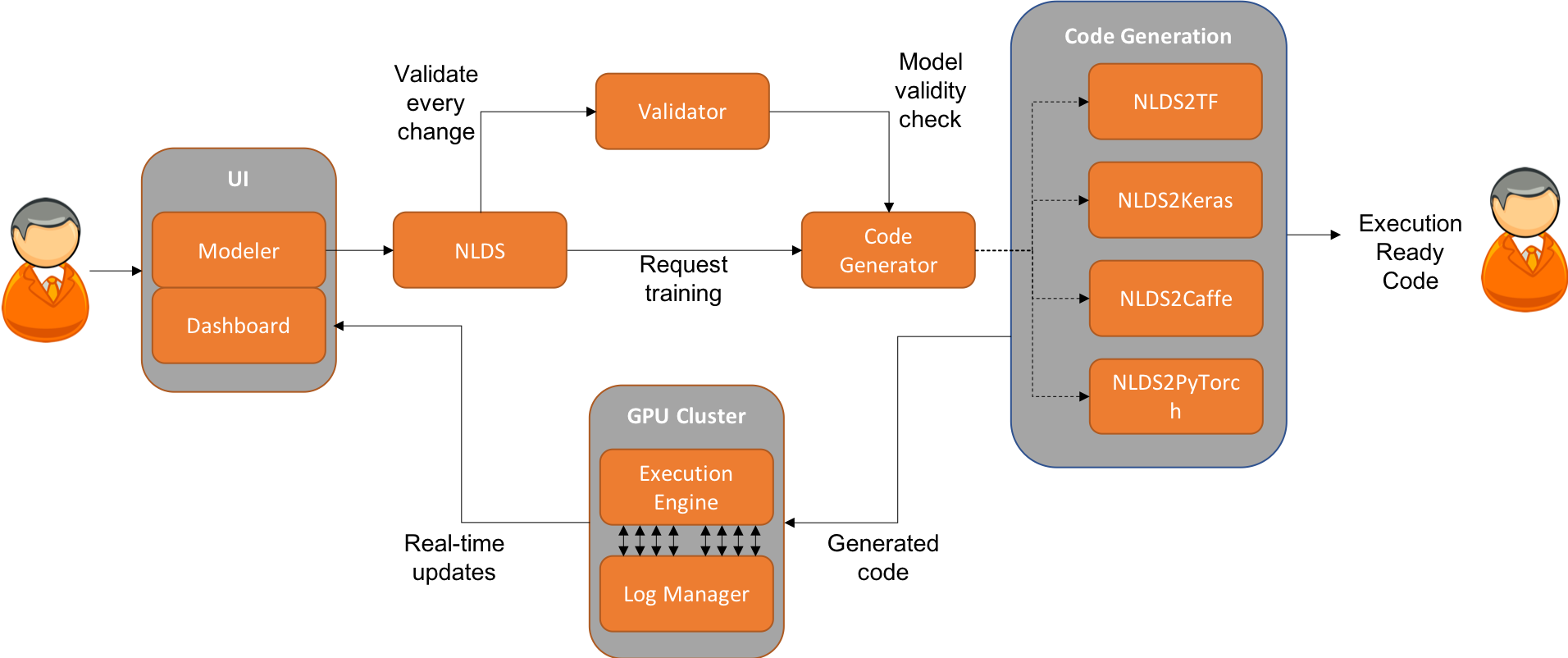}
\caption{The overall system architecture and the  components of the proposed deep learning IDE, DL-IDE.}
\label{fig:sys_arch}
\end{figure*}

\begin{figure*}
\centering
\includegraphics[width=\textwidth]{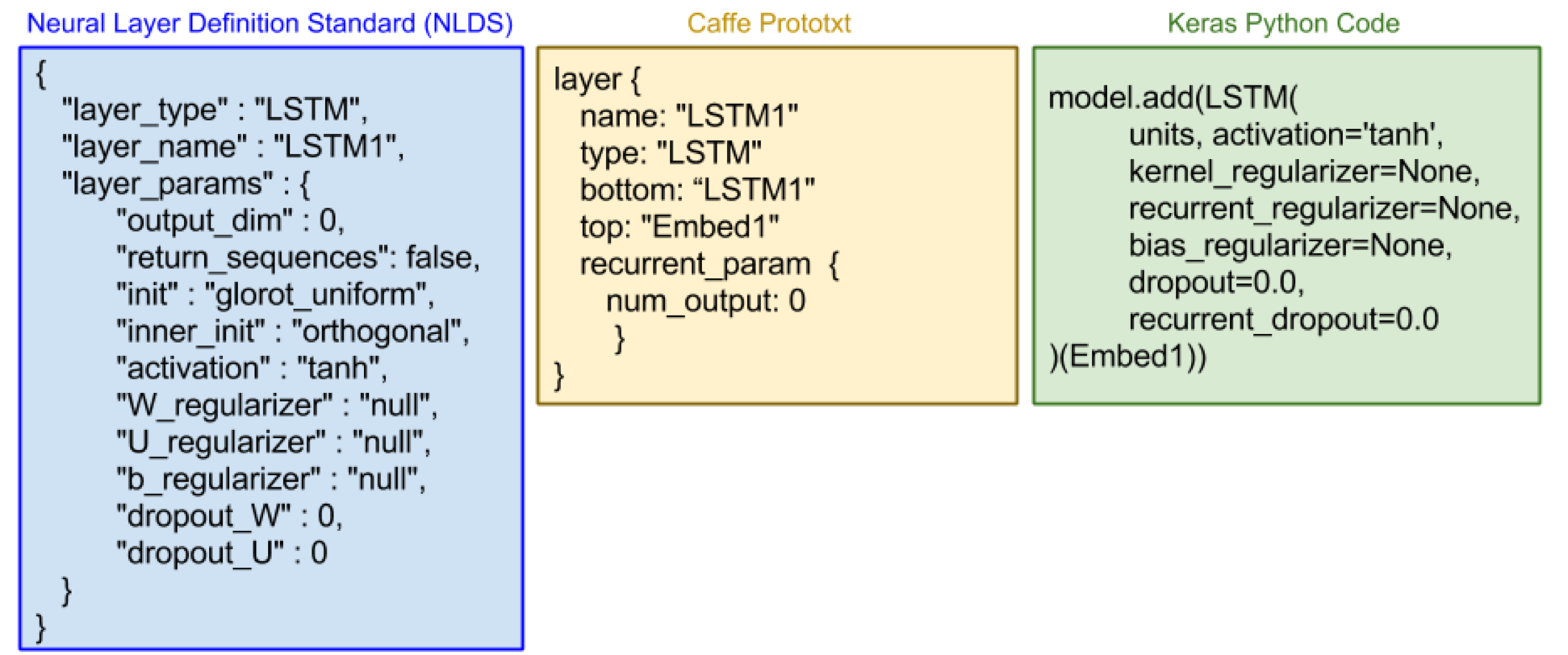}
\caption{An example of the Neural Layer Definition Standard (NLDS) for the popular LSTM layer, along with their corresponding Caffe Prototxt and Keras python code.}
\label{fig:sys_arch_nlds}
\end{figure*}

Software developers implement deep learning models by following the DL architecture details documented in research papers, blogs among others. Commonly, these deep learning models and its parameters are expressed either as a flow diagram or in a tabular format, as shown in Figure~\ref{fig:figure} and \ref{fig:table}. These architectural information are transformed into a program (code) by the developers. Currently, this coding task is time taking, laborious, and requires expert level skill~\cite{Sankaran:2017}. Motivated from these challenges, our aim is to design an IDE for deep learning where models are implemented in the same way as it is represented in documents and research papers; visually. The design goals of the proposed DL-IDE would be as follows:
\begin{enumerate}
    \item A visual ``no-code" designer to reduce the entry level barrier for DL programming.
    \item A drag-and-drop interface to design DL models as a combination of nodes (layers) and connections, corresponding to visual designs represented in research papers.
    \item A real-time validation service to verify the designed model and provide specific feedback for easy debugging
    \item An abstract DL-library independent representation of deep learning model, to enable interoperability across libraries and platforms.
    \item Export executable code from the visually constructed DL model in the library of user's choice. 
\end{enumerate}

The DL-IDE system, shown in Figure~\ref{fig:sys_arch},  consist of four main components: (i) visual drag-and-drop designer, (ii) abstract intermediate representation generator, (iii) a real-time model design validator, and (iv) code generator in multiple libraries.

\begin{figure*}
\centering
\includegraphics[width=\textwidth]{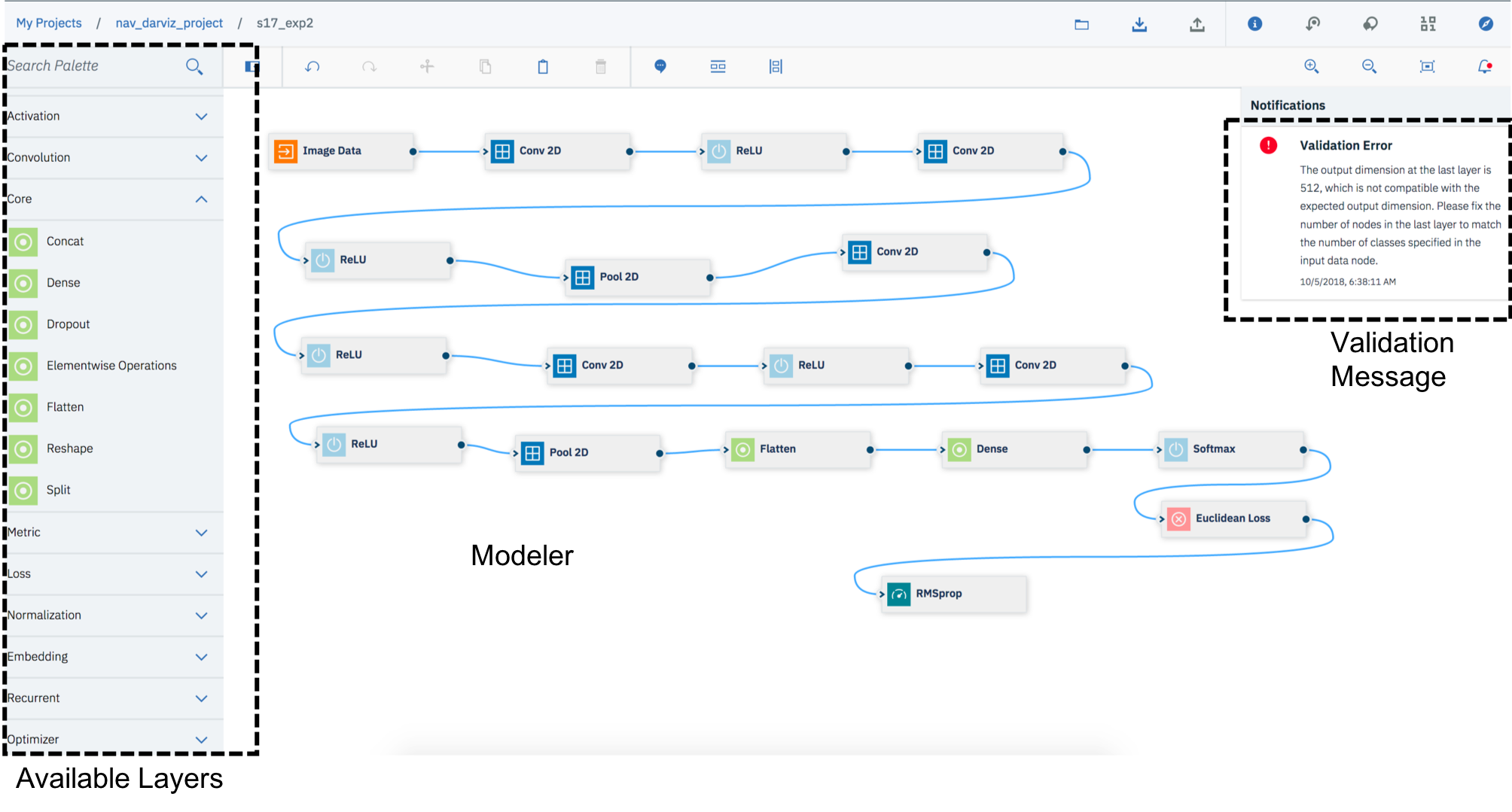}
\caption{The visual drag-and-drop canvas user interface of the DL-IDE.}
\label{fig:nnm-canvas}
\end{figure*}

\subsection{Visual Designer}
User interacts with our system through the designer component, which consists of two sub-components.  The ``DL-modeler" sub-component provides a drag-and-drop canvas for the user to author a DL model. It also provides a palette of available layers on the left and a section for displaying a detailed documentation on the right as shown in Figure~\ref{fig:nnm-canvas}. This palette groups layers performing similar function to ease development of a model for a new user.   A user can drag drag layers, configure its properties, and connect them to create a deep neural network in a language/library agnostic manner. When a user selects a layer from the palette, the system provides an initial default configuration of the parameters, which user can update based on the design needs. The system performs real-time validation of the model being designed and displays messages/errors and potential solutions on the right side section of the ``DL-modeler". The ``Dashboard" sub-component provides a real-time view of the model's training progress. The user can choose the metrics to be plotted as a graph in the dashboard, view various parameters, and hyper-parameters of the training experiment along with visualization of training data.

\subsection{Neural Layer Definition Standard}
A deep learning model is a sequential or non-sequential collection of different transformations performed on the input data to obtain the output. Each transformation, called as a layer, is the fundamental building block of a deep learning model. Neural Language Definition Standard (NLDS) is a standard, abstract representation of deep learning model that is agnostic of the underlying library. The NLDS definition of a deep neural network is a JSON object contains three main parts:
\begin{itemize}
\item \textbf{Layer definitions:} defines the parameters of every layer in the network.
\item \textbf{Layer links:} defines how the layers are connected to each other.
\item \textbf{Hyperparameter definitions:} defines model level parameters which define some characteristic properties of the model as a whole.
\end{itemize}

These three parts of the NLDS captures all aspects of a model's design. Individual library transformers can be written on top of the NLDS definition to import and export execution ready code across multiple libraries. For example, a tensorflow code could be parsed to extract the NLDS definition of the DL model, using which a PyTorch code can be generated. Thus, a lossless conversion could be performed across libraries through NLDS, as a model driven development strategy. An example NLDS for the popular LSTM layer and its corresponding Caffe Prototxt and Keras python code is shown in Fig.~\ref{fig:sys_arch_nlds}.



\subsection{DL Static Design Validation}
The validation component, validates the model being constructed in the ``DL-modeler". For every update of the DL model design in the drag-and-drop interface, the validation component checks the correctness of the overall model. Any warnings or errors are displayed as a pop-up on the validation section of the user interface. User cannot progress to the next stages of code download or training, if the constructed model is not successfully validated. Validation of a DL design happens in five levels: (1) basic NLDS structure validation, (2) layer level parameter validation, (3) static design validation for ensuring valid next and previous layers for every layer, (4) flow design validation, ensuring the data flow through the model, and (5) platform (target library) specific validation. For example, if a \textit{Convolution2D} layer is added without providing the number of filters, the validation component would prompt to provide the compulsory parameters. If a \textit{Dense} layer is added after \textit{Convolution2D} layer, the validation component would prompt that a Dense could not be added after a Convolution 2D layer, and also recommend a solution of adding a \textit{flatten} layer in between.

\subsection{Multiple Library Code Generation}
Once the constructed model is validated, the user can either opt to generate and download the source code of the model in a desired library or push the model for training. In the first case, the source code for the model is generated in the library of user's choice for download. This gives the flexibility to the user to use this tool for authoring the model in their preferred library of choice and train them in the environment of their choosing. In the second case, the model is converted into the preferred library's code and pushed to the default back-end GPU cluster for training. Currently, the tool support three libraries namely TensorFlow, Keras, Pytorch for code generation.

The GPU cluster can be considered as a pluggable extension to the IDE and is not one of its integral part as it can be replaced by any external cloud GPU service such as AWS\footnote{\url{https://aws.amazon.com/}} or Google Cloud\footnote{\url{https://cloud.google.com/}}. The main purpose of this component is to train the model on the data provided by the user. The ``execution engine" of this component contains GPU hardware and wrappers for low-level firmware like CUDA which come out-of-the-box. 
The ``log manager" is a custom component which continuously monitors the training logs for certain metrics identified by the user. These metric values are streamed to the UI dashboard and plotted as graphs. After the training is completed, the trained model is deployed and an API is exposed to the user for inference.

%% file: 4_user_study.tex
\section{Designing User Study}
To study the usability and the efficiency of the proposed visual drag-and-drop programming paradigm for DL, we conducted a user study to compare the proposed system with traditional programming paradigm and other alternative visual programming paradigms. 

\begin{figure*}
    \includegraphics[width=\textwidth]{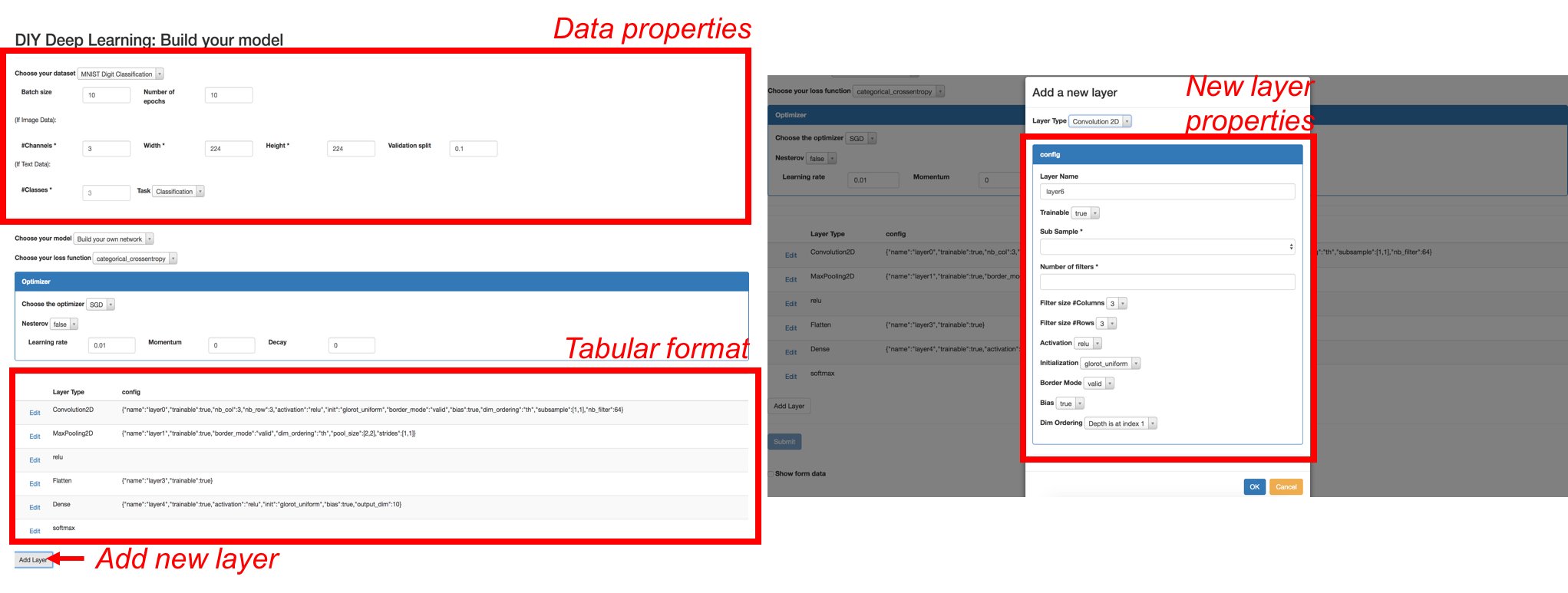}
    \caption{The tabular representation based UI system to design deep learning models.}
    \label{fig:proposed_table}
\end{figure*}

\subsection{Different Programming Paradigms}
It can be observed from Figure~\ref{fig:table} that a DL model's design could be represented in a tabular format, as well. Hence, it is worthwhile to study whether a visual tabular interface is efficient and easy for authoring DL models. To compare DL-IDE with the tabular representation of designing deep learning models, a visual tabular interface is designed as shown in Figure~\ref{fig:proposed_table}. Overall in the user study, the participants were given three different views to implement deep learning models:
\begin{enumerate}
    \item \textbf{View \#1 - Code:} Traditional coding method for implementing deep learning models. As a common standard, Microsoft Visual Studio Code~\footnote{\url{https://code.visualstudio.com/}} is used as the Python IDE. The participants were allowed to use the DL library that they are most fluent with, among the following: Keras, Tensorflow, PyTorch, Caffe.
    \item \textbf{View \#2 - Tabular:} Tabular way of designing a deep learning model, as shown in Figure~\ref{fig:proposed_table}.
    \item \textbf{View \#3 - DL-IDE (drag-and-drop):} The proposed system to design a deep learning model like a graph structure using a drag-and-drop interface, as shown in Figure~\ref{fig:nnm-canvas}.
\end{enumerate}
To avoid any bias, the participants had no prior on the motive of the experiments or on the view of interest.

\subsection{Participants} 
The user study was conducted among $18$ participants who are students from different universities. All the participants are required to have basic knowledge of developing deep learning models, in addition, to being efficient in python programming. Further, none of the participants had any prior affiliation with any of the authors and the purpose of the study remained anonymous. 

The participants filled an qualitative survey providing demographic details and their technical competence along with DL preferences. The participants comprised of  $11$ male and $7$ female participants whose age range is between $19-24$. Furthermore, the participants were highly comfortable with Python programming language and $15$ participants rated themselves as $3$ or more out of $5$ in deep learning implementation.  


To better characterise the outcome of our study, the $18$ participants are divided either as expert or intermediate based on their answers to the following eight questions from the survey:
\begin{enumerate}[label=(\Alph*)]
 \item Are you aware of Python language ? Ans: Yes/No
 \item How would you rate your comfort in Python? Scale: 1-5 
\item How would would rate your comfort in implementing Deep Learning algorithms? Scale: 1-5
\item Have you designed Deep Learning models as part of a project and/or paper ? Ans: Yes/No 
\item How many deep learning models have you designed?  
\item How comfortable are you in designing Deep Learning? Scale: 1-5 
\item Have you come across Deep Learning model in research publications (papers)? An: Yes/No 
\item How many deep learning model designs have you come across in research publications (papers) ?
\end{enumerate}

An expertise score, S, is calculated for every user using all the above eight information, with the equation below,
\begin{equation}
  \text{Expertise Score}, S = A*B+ D*(C+F)*E+ G*H
\end{equation}

The median of the expertise score is computed and all those participants above the median score are labeled experts while those below are labeled intermediate. The expertise score ranged between $19-264$ with a median of $36$. There were in total $9$ expert and $9$ intermediate users in our study.



\subsection{Experimental Procedure} 
The experiment involves participating users to implement three deep learning model designs. Details of these models are listed below as three tasks:
\begin{enumerate}
    \item \textbf{Task \#1:} Implementing a $13$ layer deep convolutional neural network (CNN) model on ImageNet~\cite{imagenet} image dataset with Conv2D, Pool2D, TanH, ReLU, Flatten, Dense, and Softmax as the set of unique layers. 
    \item \textbf{Task \#2:} Implementing a $16$ layer deep convolutional neural network (CNN) model on CIFAR-10~\cite{cifar} image dataset with Conv2D, Pool2D, ReLU, Flatten, Dense, and Softmax as the set of unique layers.
    \item \textbf{Task \#3:} Implementing a $6$ layer deep recurrent neural network (RNN) model on text classification dataset with Embedding, LSTM, Dense, and Softmax as the set of unique layers.
\end{enumerate}
To avoid any known model bias, popular deep learning models were ignored for this study and the above three novel models were designed from scratch. Based on the author's experience in deep learning, they are also of equal difficulty level in implementation.

Each participant was asked to perform three experiments: \textbf{Experiment 1}, \textbf{Experiment 2}, and \textbf{Experiment 3} where, in each experiment, they implemented one of the three tasks in one of the three views. To counter for the learning effect, the task-view combination for all $18$ participants were randomly chosen upfront. Overall, $54$ experiments were conducted among $18$ participants. 

A time limit of $15$ minutes was provided to perform each experiment, and the participants were given the choice to continue or give up if it goes beyond $15$ minutes. An instruction sheet with detailed verbose instructions of the implementation task was provided in a written form before the start the task. A demo video explaining the features of each view was played to the participant just before performing the corresponding task. Additionally, users had access to internet throughout the experiment, and there were no restriction on referring to sources in the internet to complete the task.  Throughout the task, one interviewer stayed in the room to help with infrastructural or operational needs, however she did not indulge in any technical discussion with the participant. All participants were given the same laptop and infrastructure to complete the task. Each study lasted around $60$ minutes and the participants were provided with refreshments. To increase the competitive nature of the tasks, the top three participants were promised a reward of $20$ USD. The instructions sheets and videos, the qualitative survery results and the task-view combination of all the 18 participants is available here \url{https://dl-ide.github.io/}.

At the end of each experiment, two different surveys were conducted: (1) System Usability Scale (SUS)~\cite{brooke1996sus} to study for usability of the corresponding view, and (2) NASA Task Load Index (TLX)~\cite{hart1988development} to study the experimental task load. Also, the time taken for the participant to complete a particular experiment is noted along with the entire screen recording of the activities.


%% file: 5_study_results.tex
\begin{figure}[!b]
    \includegraphics[width=0.5\textwidth]{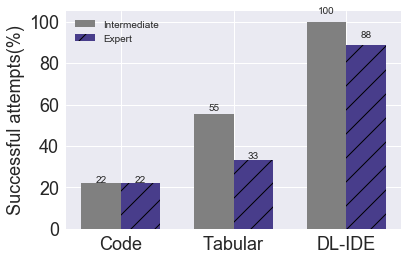}
    \caption{Plot showing the successful attempts by $9$ expert and $9$ intermediate participants in each of the three views: code, tabular, and DL-IDE.}
    \label{fig:sucess_views}
\end{figure}

\section{Results of User Study}

The results of study are analyzed and summarized in this section.

\subsection{Analysis of Accuracy and Time Taken}
\begin{figure*}
    \begin{subfigure}[b]{0.5\textwidth}
        \includegraphics[width=\textwidth]{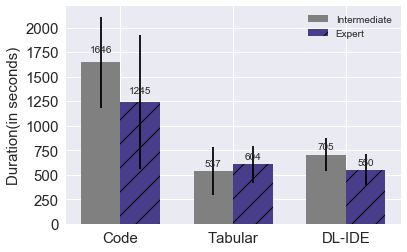}
        \caption{All experiments}
        \label{fig:time_platform}
    \end{subfigure}
    \begin{subfigure}[b]{0.5\textwidth}
        \includegraphics[width=\textwidth]{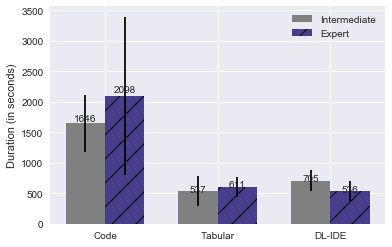}
        \caption{Only successful experiments}
        \label{fig:sucess_platform}
    \end{subfigure}
\caption{Plot showing the mean time duration by all users in doing in the experiments across the three different views: code, tabular, DL-IDE. The restricted time provided to the participants is $15$ minutes ($900$ seconds) though they were later given a choice to take additional infinite time to finish the task.}
\end{figure*}

\begin{figure}
    \begin{subfigure}[b]{0.4\textwidth}
        \includegraphics[width=\textwidth]{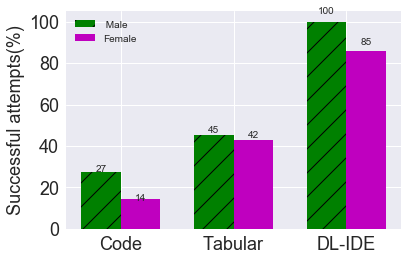}
\caption{Successful experiments}
\label{fig:gender_sucess_platform}
    \end{subfigure}
    \begin{subfigure}[b]{0.4\textwidth}
        \includegraphics[width=\textwidth]{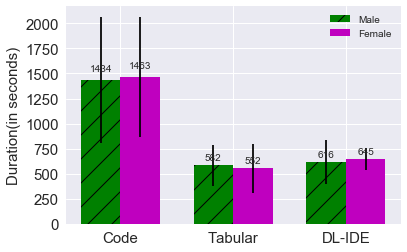}
\caption{Mean time duration}
\label{fig:gender_time}
\end{subfigure}
\caption{Plot showing the successful attempts and the mean time taken by $11$ male and $7$ female participants to perform experiments in three different views: code, tabular, DL-IDE.}
\end{figure}

\begin{figure}
 \begin{subfigure}[b]{0.4\textwidth}
        \includegraphics[width=\textwidth]{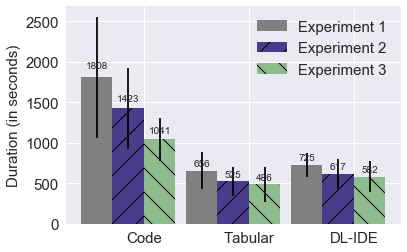}
        \caption{All experiments}
        \label{fig:time_exp_platform}
    \end{subfigure}
     \begin{subfigure}[b]{0.4\textwidth}
        \includegraphics[width=\textwidth]{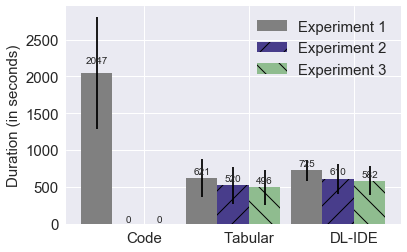}
        \caption{Only successful experiments}
        \label{fig:time_exp_platform_success}
    \end{subfigure}
    \caption{Plots showing the mean time duration taken by the $18$ participants in doing the tasks in experiment 1, 2, 3 in all the three views.}
\end{figure}

\begin{figure}[!t]
\centering
\includegraphics[scale=0.5]{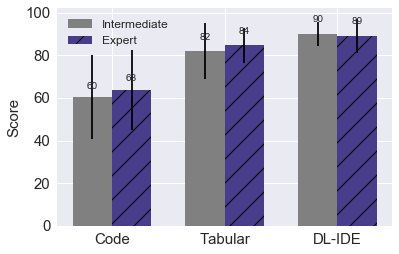}
\caption{Mean SUS score obtained for the three different views: code, tabular, and DL-IDE. }
\label{fig:sus_score}
\end{figure}

Each of the $9$ expert and $9$ intermediate participants performed one experiment in each of the three views: code, tabular, and DL-IDE. We analzye the responses and results by answering 5 research questions.

 \textbf{RQ1:} ``\textit{Does the visual drag-and-drop view enable users to implement the deep learning models more accurately?}". The accuracy of the task was verified by executing the implemented deep learning model from each of the three views.

As observed from Figure~\ref{fig:sucess_views}, $17$ of the $18$ participants were able to finish the task successfully using the DL-IDE view, while only $4$ participants were able to implement the task using the code view (traditional programming) and only $8$ participants were successful with the tabular view. Note that this result is despite providing infinite time beyond the instructed $15$ minutes to finish the experiment. Thus, it is safe to conclude that almost $78\%$ ($14/18$) participants gave up without being able to complete the deep learning model implementation using coding. Out of these $18$ participants, $11$ coded using Keras, $5$ using Tensorflow, $2$ used PyTorch. Irrespective of the choice of the implementation library, it can be observed that programming view remained a challenge.

Another reason for the higher success rate in DL-IDE can be attributed to the static validation component developed as a part of our system. Manually observing the screen recordings of the experiments, we noted that more often participants found it highly challenging to debug the error in their implemented code and that majorly influenced them to give up. The validation component provided easy guidance with precise message and highlighted specific layers for the users to fix the bugs, improving the success rate for implementing deep learning models in DL-IDE.

Also, preference for DL-IDE is supported by the qualitative survey conducted prior to the experiments where the participants where asked the question, ``When reading in research papers, what style of deep learning model representation would you prefer?" with choices as (a) Tabular rows and columns, or (b) Flow Figure as an Image, or (c) others. $15$ of the $18$ participants preferred deep learning models being represented as a ``Flow Figure" which is the representation in DL-IDE.

\textbf{RQ2:} ``\textit{Can deep learning models be implemented much faster using the visual drag-and-drop view?}".

Figure~\ref{fig:time_platform} depicts the mean and the standard deviation for time duration, spent by participants in performing the experiments across different views. Figure~\ref{fig:sucess_platform} depicts the mean time taken and the standard deviation of only those experiments which were successfully completed by the participants. Irrespective of the success outcome of the experiment, the mean time taken for coding is much more than the visual methods of designing deep learning models. The mean time taken for implementing deep leaning models using either of visual programming view is in the range of $8.9-11.8$ minutes ($536-705$ seconds) while for the traditional programming view the mean time taken is $27.4-35$ minutes ($1646-2098$ seconds). Though the mean time taken in tabular view is similar to DL-IDE, the success rate of designing accurate deep learning designs in the DL-IDE view is much higher (more than $70\%$) as debugging model designs is much more intuitive and natural.

 \textbf{RQ3:} ``\textit{Is DL-IDE suited well for both intermediate and expert users in implementing deep learning models?}".

From Figure~\ref{fig:time_platform}, it can be observed that the mean time duration to implement deep learning models for intermediate participants ($1646$s) is much higher than for expert users ($1245$s), for code view. However, using DL-IDE the difference in mean time duration is reduced to $705$s for intermediate users and $560$s for expert users.
It can also be observed that in DL-IDE, the variance of time taken is comparable irrespective of user expertise, showing that DL-IDE caters to all the users equally irrespective of their prior expertise in deep learning. 

\textbf{RQ4:} ``\textit{Does participant's gender affect the outcome of the experiments?}".

Figure~\ref{fig:gender_sucess_platform} and Figure~\ref{fig:gender_time} capture the successful attempts and the mean time duration in performing the experiments across gender. The mean time spent by all the participants is similar in all the views. Also, gender does not act as a biasing factor in successful completion of the experiment as there is no significant difference across the gender.

\textbf{RQ5:} ``\textit{Is there a learning bias displayed by the participants based on the sequence in which tasks are performed?}".

Figure~\ref{fig:time_exp_platform} and Figure~\ref{fig:time_exp_platform_success} attempts to address an important question of whether the sequence in which the views are presented to the participant affect the mean time to complete the experiment. Mean time across all the experiments is highly similar for DL-IDE and tabular views demonstrating that the order does not affect the performance. However, it can be observed in Figure~\ref{fig:time_exp_platform} that the mean coding time is higher if it is presented as Experiment 1 compared to when it is presented as Experiment 2 and Experiment 3. This can be attributed to the human tendency to spend more time in completing the task for the initial experiments than for the later experiments, where there is a tendency to give up in lesser time. Figure~\ref{fig:time_exp_platform_success} shows that even in successful experiments, the order does not affect the time performance in DL-IDE and tabular views. For coding view, there were no successful attempts in Experiment 2 and Experiment 3. 

\subsection{System Usability Scale (SUS)}
To study the usability of the three views, System Usability Scale (SUS)~\cite{brooke1996sus} was used. 
It consists of a set of $10$ questions with every alternate question having a positive and negative intent. Each question has five different choices ranging from strongly agree to strongly disagree. SUS is known to work with smaller sample sizes, as well. A SUS score of $68$ is considered average and any system with a SUS score of above $68$ is considered usable. 
The SUS questionnaire was conducted for every participant at the end of every experiment.
Figure~\ref{fig:sus_score} indicates the mean SUS score computed for the different views based on the user expertise. It can be observed that the coding view has an average SUS score of $60$  for the intermediate participants and $68$ for the experts based on their feedback. Thus, even for experts the usability of the system is just average. However, the average SUS score for DL-IDE is $90$ and $89$ for intermediate and experts respectively, the highest in our studies. From our study, we can infer that the drag-and-drop interface in DL-IDE can be considered as the   highly usable system for both intermediate and experts, alike.

\subsection{NASA TLX Results}

\begin{figure}[!t]
        \includegraphics[scale=0.5]{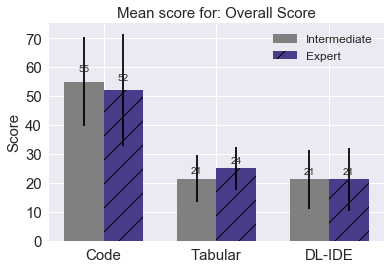}
\caption{Plots describing the overall mean NASA TLX score as well as the NASA TLX score for each of the six dimensions for all the three views.}
\label{fig:nasa_tlx}
\end{figure}

NASA Task Load Index (TLX)~\cite{hart1988development} is a popular tool used to measure the perceived workload of a user after performing a specific task. The questionnaire is multidimensional, measuring the workload impact across six different dimensions. The participants were provided with the NASA TLX 
The plot in Figure~\ref{fig:nasa_tlx} captures the overall workload faced by the participants averaged along the six different dimensions. From the results obtained on the individual dimensions, we observe that the mental demand such as looking, calculating, and searching required by the code view is significantly high as users often have to refer documentation or check syntax or perform layer size calculation. DL-IDE view has least demand for mental strength as it provides easy to use interface for fundamental deep learning functionalities. The physical demands such as typing, scrolling, and clicking are less required in DL-IDE and tabular view as compared to code view. Both the visual programming views have lesser physical demand, as they don't involve much of typing and window switching. Both the visual programming interfaces require less temporal demand, as the time to create or implement layers and connect them in these views are less. Further, participants were not satisfied with their performance in code view and tabular view. The key reason of high dissatisfaction is that they could not complete the experiments in given time which we observed from them as feedback post experiments. Effort required to design a deep learning model in DL-IDE is less compared to other views, as DL-IDE provides easy to use drag-and-drop interface to create the deep learning models. The study also illustrates the frustration of the participants while performing the experiments, where less effort and less demand of mental and physical activity leads users to less frustration. The mean workload follows the same footprint as other workload figures and based on overall score we can conclude code view has the highest workload than the tabular view or DL-IDE view.

%% file: 7_design_implication.tex
\section{Design Implications}

The results obtained in the user study are in favour of the proposed DL-IDE in terms of successful completion of tasks ($94\%$), time required to perform a task ($9-12$ minutes), system usability (SUS score of $90$), and least workload (NASA TLX of $21$). However, the obtained results may not be generalizable and the design study had a few implications, summarized as follows:

\begin{enumerate}
    \item The user study is conducted using a set of $18$ participants from a restricted demography and age group.
    \item The participants are students selected at random from various universities. The prerequisites for a student to participate in the user study were proficiency in python and minimum development knowledge in deep learning. Students who had done a couple of projects in deep learning turned out to be experts and those who have done only one or two projects turned out to be intermediate in our group of participants. However, researchers in deep learning could be the actual set of experts and studies have to be conducted using them to verify if similar results could be extrapolated. With the existing user studies, it is safe to conclude that a visual drag-and-drop based DL-IDE aids 
    users in better implementing deep learning models. 
    \item The primary purpose of this research paper is to methodically study and compare visual programming paradigm (drag-and-drop) vs. traditional programming paradigm in deep learning implementation. DL-IDE is one of the few drag-and-drop frameworks in the literature for deep learning implementation, with enhanced features such as (1) Neural Language Definition Standard (NLDS), (2) static design validation service, and (3) multiple library code generation. Comparing different drag-and-drop frameworks for implementing deep learning models is out of scope for this research work and is considered as a future study.
\end{enumerate}

On the positive side, the results obtained from the studies conducted in this research has some strong, impacting implications on the research and development in the deep learning community. The major implications are summarized below:
\begin{enumerate}
    \item Owing to the graphical structure of deep learning models, it is established that a visual drag-and-drop framework could be an efficient platform for implementing deep learning models compared to the traditional programming paradigm. This encourages the developers of libraries such as Keras, Tensorflow, PyTorch to align and support visual programming frameworks.
    \item For students and beginners, the drag-and-drop framework provides an easy, smooth, and confident route to adapt deep learning with very low initial learning curve. Additionally, such visual tools could be used as effective teaching tools in classrooms to intuitively explain the design choices in a deep learning model.
    \item With the growing requirement of having deep learning as a skill for software engineers, industries are looking at upskilling hundreds (may be thousands) of their existing software engineers and data scientists. DL-IDE could enable this transition at industry level, democratizing deep learning implementation for novice software engineers.
    \item Existing implementation of research papers and algorithms are mostly available in only a single library offering very little scope for interoperability into other libraries. With the aid of DL-IDE and abstract NLDS representation, standardization of deep learning models could be obtained. Research papers could make available their implementation in standard NLDS format and different users could generate the same implementation in different libraries of their choice. Furthermore, the flow figure used in research papers could be standardized as the visual drag-and-drop screenshot to make research papers more uniform and easy to read.
\end{enumerate}

%% file: 8_conclusion.tex
\section{Conclusion and Future Works}
In the recent years, deep learning has been the driving force for most of the popular AI applications such as speech recognition, image processing, and natural language processing. Therefore, a lot of aspiring developers, students, and enterprises are wanting to solve different problems using deep learning. However, the democratization of deep learning is not truly achieved and remains a tool for the experts because of the (1) high initial learning curve, (2) lack of interaction across existing deep learning libraries, and (3) need for theoretical deep learning knowledge.

To address these challenges, in this research work we presented DL-IDE, an intuitive drag-and-drop based user interface for visually constructing deep learning models. The tool provides an end-to-end development capability including model designing from scratch, model design validation, code generation in multiple libraries, model training to deployment, and model inference. Further, we validated the usability of our system by conducting comprehensive experiments with $18$ participants on three different tasks of equal difficulty using three different set of tools (views). The participants were able to successfully complete the implementation task $94\%$ of the times using DL-IDE, however, only $22\%$ of the times using the traditional programming method with any library of their choice. On an average, participants took about $9-12$ minutes to implement the tasks using DL-IDE, while successful participants took more than $27$ minutes to implement the tasks using programming. DL-IDE obtained a System Usability Scale (SUS) score of $90$ compared to a score of $68$ for programming, establishing the ease and usability of DL-IDE. DL-IDE obtained the least NASA-TLX workload index of $21$ compared with a index value of $55$ for traditional programming.

As there exists multiple drag-and-drop frameworks in the literature, it would be interesting to compare and study the multiple frameworks in terms of efficiency and usability. Also, user studies could be conducted on a more diverse set of participants and experts to obtain more generalizable conclusions.